%
%
%
%
\input mtexsis.tex

\paper
\tenpoint
\input epsf.tex
\superrefsfalse
\referencelist
\reference{bender} {Bender R., Saglia R. \& Gerhard O.} 1994,
{\journal  Monthly Notices of the Royal Astronomical Society; 269, 785-789(1994)} \endreference
\reference{BT} {Binney, J. and S. Tremaine, 1987}{ \it Galactic
Dynamics}{ } \endreference
 \reference{fisher} {Fisher, D., et al. } {\sl ``The dynamics and
     structure of the S0 galaxy NGC7332''} {\journal Astronomical Journal;107,
     160-172(1994)}\endreference
\reference{ Dejonghe} Vauterin, P.; Dejonghe, H.
 {\sl `` A method for solving the linearized Boltzmann
                    equation for almost uniformly rotating stellar
                    disks.} {\journal  Astronomy and Astrophysics;296,380 (1995)}\endreference
\reference{Fricke} {Fricke, W.} {} {\journal Astrnachr;280,
193-(1952)}\endreference
\reference{Evans2} {Evans, N. and P. de Zeeuw} {\sl"Potential-density
Pairs for Flat Galaxies."} {\journal Monthly Notices of the Royal Astronomical Society;257,
152-176(1992)}\endreference
\reference{Evans} {Evans, N. } {\sl"The Powerlaw disks and halos."} {\journal Monthly Notices of the Royal Astronomical Society;257,
152-176(1993)}\endreference
\reference{Hunter} {Hunter, C.} {\sl"Instabilities of stellar discs."}
{in {\sl Astrophysical Disks} ed. S.F. Dermott, J.H. Hunter,
Jr, R.E. Wilson, Annals of the New York Academy of Sciences,
Volume 675, pp 22-30 (1992)}\endreference
\reference{Qian} {Hunter, C. \& Qian, E.} {\sl" 
Two-integral distribution functions for axisymmetric galaxies."}
{\journal  Monthly Notices of the Royal Astronomical Society;262,
401-428(1993)}
\endreference
\reference{kalnajs71} {Kalnajs, A.} {\sl"Dynamics of Flat Galaxies. I."}
{\journal Astrophysical Journal;166, 275-293(1971)}\endreference
\reference{kalnajs76} {Kalnajs, A.} {\sl"Dynamics of Flat Galaxies II,
Biorthonormal Surface density potential pairs for finite disks."}
{\journal Astrophysical Journal;205, 745-761(1976)}\endreference
\reference{DLB60} {Lynden-Bell, D.} {\sl"Stellar Dynamics: Exact
solutions of the self-gravitating equations"} {\journal Monthly Notices of the Royal Astronomical Society;123, 447-458(1960)}\endreference
\reference{Mestel}{Mestel, L.} {\sl ``On the galactic law of
rotation.''} \journal Monthly Notices of the Royal Astronomical Society;126,553-575(1963)\endreference
\reference{Miyamoto}{Miyamoto}{\sl ``A class of disk-like models 
for self-gravitating stellar systems''} \journal Astronomy \& Astrophysics;30,441-454(1974)\endreference
\reference{Cannon} {Pichon, C. \& Cannon, R.} {\sl"Numerical linear 
stability for round galactic disks."} {to be submitted to  the Monthly Notices of the Royal
 Astronomical Society (1996)}\endreference
\reference{PLB2} {Pichon, C. \& Lynden-Bell, D.} {\sl"Numerical linear 
stability for round galactic disks."} {\journal MNRAS;280,1007-1026.
 (1996)}\endreference

\reference{Sawamura} {Sawamura, M.} {\sl"A Normal Mode Analysis of a
Class of Self Gravitating Stellar Discs with Differential Rotations."}
{\journal PASJ;490, 279-311(1988)}\endreference
\reference{Toomre63}{Toomre, A.} \journal Astrophysical Journal;138,385-394(1963)\endreference
\endreferencelist

\title
Equilibria of flat and round galactic disks.
\endtitle
\author 
C. Pichon$^{1,2}$  and D. Lynden-Bell$^{1}$
$1$: Institute of Astronomy, The observatories, Cambridge CB3OHA, UK

$2$: CITA, 60 St. George Street, Toronto, Ontario M5S 1A7, Canada.
\endauthor
\abstract
	A general method is presented for constructing distribution functions
for flat systems whose surface density and Toomre's Q number
profile is given.  The purpose of these functions is to provide
plausible galactic models and assess their critical stability with
respect to global non axi-symmetric modes.  The derivation  may be
carried out for an azimuthal velocity distribution (or a given specific
energy distribution) which may either be observed or chosen
to match a specified temperature profile. 
 Distribution functions describing stable models with
realistic velocity distributions for power law disks, the Isochrone
and the Kuzmin disks are provided.
 Specially simple inversion formulae are also given for finding
distribution functions for flat systems whose surface densities are
known.
\endabstract
\vskip 0.5cm
\centerline{{\it  accepted for publication by the } 
 Monthly Notices of the Royal Astronomical Society}



\section{Introduction}  
\enddoublecolumns

Over the next decade, ample and accurate observational data on the
detailed kinematics of nearby disk galaxies will become available. It
will be of great interest to link these observations with theoretical
models for the underlying dynamics. 
The problem of finding the distribution function for an axially
symmetrical system may formally be solved by using Laplace transforms
(Lynden-Bell 1960\cite{DLB60}) or by using power series (Fricke
1952\cite{Fricke}, or more recently Vauterin 1995\cite{Dejonghe}). 
 Kalnajs (1976)\cite{kalnajs76}, followed  by
Miyamoto (1974)\cite{Miyamoto}
chose specific forms of distribution function because the flat
problem has no unique solution.
Distribution functions for the power-law disks  were also  recently derived
by Evans (1993)\cite{Evans} while  Hunter \& Qian (1993)\cite{Qian} presented
an inversion scheme which can be applied for thickened disks with two integrals.
For flat disks, it is desirable to use the functional freedom left in $f$ in order to
construct a distribution function which accounts for all the
kinematics, either observed or desired (i.e. which accounts for the
line profiles observed, or which are marginally stable to radial
modes).
 Independent measurement of the observed
radial and azimuthal velocity distribution functions could, for
instance, be contrasted with predictions arising from the
gravitational nature of the interaction.  Indeed the laws of the
motion and the associated conserved quantities together with the
assumption that the system is stationary put strong constraints on the
possible velocity distributions.  This is formally expressed by the
existence of an underlying distribution function which characterises
 the dynamics completely. The determination of realistic distribution
functions which could account for observed line profiles is therefore
an important project vis a vis the understanding of galactic
structure.  Producing theoretical models that accounts globally for the
observed line profiles of a given galaxy provides a unique opportunity
to inspect the current understanding of the dynamics of S0
galaxies. It should then be possible to study quantitatively all
departures from the flat axisymmetric stellar models.  Indeed,
axisymmetric distribution functions are the building blocks of all
sophisticated stability analyses, and a good phase space portrait of
the unperturbed configuration is often needed in order to asses the
stability of a given equilibrium  state.  Numerical
N-body simulations also require sets of initial conditions which should
reflect the nature of the equilibrium.

For a flat galaxy all the stellar orbits are confined to a plane and by Jeans' 
theorem the steady state mass-weighted distribution function must be of the 
form $f = f(\varepsilon, h)$, where the specific energy, $\varepsilon$,
and the specific angular momentum, $h$, are  given by
$$\varepsilon = {\textstyle{1\over2}} (v_R^2 + v_\phi^2) - \psi \quad
{\rm and} \quad h = R\, v_\phi . \EQN 1.1$$
The surface density, $\Sigma (R)$, arises from this distribution of stellar orbits 
provided that the integral of $f$ overall bound velocities is $\Sigma (R)$, 
\ie 
$$\Sigma (R) = \int \int f(\varepsilon, h) dv_R dv_\phi \, , \EQN 1.3$$
where the integral is over the region ${\textstyle{1\over2}} (v_R^2 + 
v_\phi^2) < \psi$.
The following inversion methods assume that
 $\Sigma (R)$ is given and its potential on the plane, 
$\psi (R)$ is known.
This is achieved either by requiring self-consistency via Poisson's equation
$$ \nabla^2 \psi = - 4 \, \pi \, \delta(z)\, \Sigma(R) \,, \EQN fish $$ or
alternatively by assuming that the disk is 
embedded in a halo and choosing both $\Sigma$ 
and $\psi$ independently.
At constant $v_\phi$ and $R$, $v_R\, dv_R = d\varepsilon$ and hence, 
using \Eq{1.1},
$$dv_R = d\varepsilon [2(\varepsilon + \psi) - h^2 R^{-2}]^{-{1/2}} \, . \EQN 
2.1$$
Furthermore, at constant $R$,  $dh = R\, dv_\phi$, and  therefore
\Eq{1.3} yields
$$\Sigma (R) = \int^{+\sqrt{2R^2\psi}}_{-\sqrt{2R^2\psi}} 
2\int^0_{{h^2\over 2R^2}-\psi} {f(\varepsilon, h)\, d\varepsilon\, 
dh\over\sqrt{2(\varepsilon+\psi) R^2-h^2}} \, . \EQN 2.2$$
The factor of 2 arises because $v_R$ takes both positive and negative values 
 which are mapped into the same range of $\varepsilon$ which depends 
only on $v_R^2$.
Similarly,
$$\Sigma (R) v_R^2  = \int^{+\sqrt{2R^2\psi}}_{-\sqrt{2R^2\psi}} 
2\int^0_{{h^2\over 2R^2}-\psi} {f(\varepsilon, h)\,
\sqrt{2(\varepsilon+\psi) R^2-h^2} \, \,  d\varepsilon\, 
dh} \, . \EQN 2.2-2$$
The basic inversion problem is concerned with finding the class of 
$f(\varepsilon, h)$ compatible with the constraints \Ep{2.2} and \Ep{2.2-2}.

Two classes of inversion methods are constructed and implemented in
this paper. These lead to classical distribution functions compatible
with a given a given surface density (section 3), or a given surface
density and a given pressure profile (section 2).  The former
technique relies on an Ansatz and yield direct and general methods for
the construction of distribution functions for the purpose of
theoretical modelling.  The latter technique yields distribution
functions that correspond to given line profiles (or rather, to the
shape of the line profiles induced by the velocity distributions
which will here loosely be called line profiles) which may either be
postulated or chosen to match the observations.  In fact, only a
sub-sample of the observation is required to construct a fully
self-consistent model which accounts, in turn, for all the observed
kinematics. The corresponding redundancy in the data may be exploited
(as sketched in section 4) to address the limitations of the
description.

\section{ Distribution functions for given kinematics.}
 This global inversion method  introduces an
intermediate observable, $F_\phi(R,v_\phi)$, the distribution function
for the number of stars which have azimuthal velocity $v_\phi$ at radius
$R$ (or alternatively $G_\varepsilon(\varepsilon,R)$ the distribution
function for the stars which have specific energy
$\varepsilon$ at radius $R$).  This technique may be applied to the
reconstruction of distribution functions accounting for {\sl observed
} line profiles. In fact, given some symmetry assumptions about the shape
of an observed galaxy, it is shown that only a subset of the available
line profiles -- say the azimuthal velocity distribution -- is
required to re-derive the complete kinematics.  Therefore this method 
provides a general procedure for constructing self-consistent models for
the complete dynamics of disk galaxies. These models predict radial
velocity distributions which may, in turn, be compared with observations
as discussed in section~4.
\nl
Alternatively, a `natural' functional form for $F_\phi$ or
  $G_\varepsilon$ may be postulated and parameterised so that it is
  compatible with imposing the surface density, the average azimuthal
  velocity and the azimuthal pressure profiles (or equivalently the
  Toomre number $Q$, as $Q$ follows from the equation of radial
  support).  The distribution function $f(\varepsilon,h)$ of this disk
  follows in turn from $F_\phi$ or $G_\varepsilon$ via simple Abel
  transforms.  This prescription is very general and especially useful
  when setting the initial conditions of numerical N-body stability analysis.
 \subsection{Inversion via line profiles}
 The number of stars which have azimuthal velocity $v_\phi$ within $d v_\phi$ 
 at radius $R$ reads
 $$F_\phi(R,v_\phi)=\int  {f(\varepsilon,h) 
 d \, v_{_R}  }  = \sqrt{2 } \int \limits_{-Y}^{0}  {f(\varepsilon,h) 
  \over \sqrt{\varepsilon + Y} } \,d  \varepsilon \, , \EQN F$$
 where the effective potential, $Y$,  is given by
 $ Y = \psi - \ h^2 \,R^{-2}/2=\psi - {v_\phi^2 / 2 } $.
The line profile  $F_\phi(R,v_\phi)$ may be expressed in terms of $(h,Y)$. 
Indeed the identity $ Y= \psi - h^2 \,R^{-2}/2 $ may be solved for 
$R$ which yields $R(h,Y)$. The azimuthal velocity  $v_\phi$ becomes in turn a function of   $h,Y$ defined
by $v_\phi = h/R(h,Y)$ for each branch corresponding to the two roots
of $ Y= \psi - h^2 \,R^{-2}/2 $ (as illustrated in \Fig{path}).
 Calling  ${\tilde F}_\phi (h,Y) \equiv F_\phi(R,v_\phi)$ 
 allows the inversion of \Eq{ F}  by an Abel transform:
 $$f(\varepsilon,h) ={1 \over \sqrt{2} \pi} \int \limits_0^{-\varepsilon} 
 {\partial {\tilde F}_\phi \over \partial Y}
   { \,d  Y \over \sqrt{ (-\varepsilon) - Y} } \, , \EQN invertAbel $$
 where the partial derivative is taken at constant $h$.  It was
assumed here that the distribution $F_\phi(R,v_\phi)$ vanishes at the
escape velocity.  Note that $\tilde F$ yields both the symmetric and
the antisymmetric parts of the distribution function.  The {\it r.h.s
} of \Eq{invertAbel} is advantageously re-expressed in terms of the
variables $(R,h)$ given that (for monotonic integration)
$$\left({\partial \tilde F_\phi \over \partial Y }\right)_h \,  dY =
 \left({\partial { F_\phi } \over \partial R}\right)_h
\left({\partial { R  } \over \partial Y}\right)_h \, d Y =
  \left({\partial { F_\phi } \over \partial R}\right)_h \,
 {\rm sign}\left[  \left({\partial { R  } \over \partial Y}\right)_h \right] \, dR \,. \EQN $$ 
This yields two contributions for \Eq{invertAbel}
$$ f(\varepsilon,h) ={1 \over \pi} \int \limits_{R_e}^{R_p} 
 {\left({\partial { F_\phi } / \partial R}\right)_h \,d  R \over \sqrt{ h^2/R^2 - 2\psi(R)-2\varepsilon }}-
{1 \over  \pi} \int \limits_{R_a}^{\infty} 
{  \left({\partial { F_\phi } / \partial R}\right)_h\,d  R \over \sqrt{ h^2/R^2 - 2\psi(R)-2
\varepsilon }} \, ,  \EQN invertAbelR $$
where $R_p(h,\varepsilon)$, and $R_a(h,\varepsilon)$  are, respectively, the apogee and  perigee of the star 
with invariants $h$ and $\varepsilon$, and $R_e(h)$ is the inner radius of a star
on a ``parabolic'' (zero energy) orbit with momentum $h$. Note that
the derivative
 in \Eq{invertAbelR} is
performed keeping $h= R \, v_\phi$ constant. The contributions to
the two integrals is illustrated on \Fig{path}. \nl
Equation \Ep{invertAbelR} yields the unique distribution function compatible with an
 observed azimuthal
velocity distribution. All macroscopic properties of the flow follow from $f$.
For instance the surface density reads
$$ \Sigma   \equiv   \int
 \left[\int  {f_+(\varepsilon,h) 
 d \, v_{_R}  } 
 \right] \, d v_\phi = 2  \int \limits_0^{v_e}
 F_\phi(R,v_\phi) \, d v_\phi \, , \EQN $$
while the radial velocity distribution, $F_R(R,v_R)$, obeys
$$ F_R(R,v_R)= 
{1\over R} \, \int\limits_{-R(2 \psi -v_R^2)^{1/2}}^{R(2 \psi -v_R^2)^{1/2}} f({v_R^2 \over 2 }+ 
{h^2 \over 2 R^2}-\psi(R) ,h)\, d h \, \, . \EQN FVR   $$
Both $F_R$ and $\Sigma$ may in turn be compared with the corresponding
observed quantities.
\figure{path}
\centerline{\hbox{\epsfysize=10cm\epsfbox{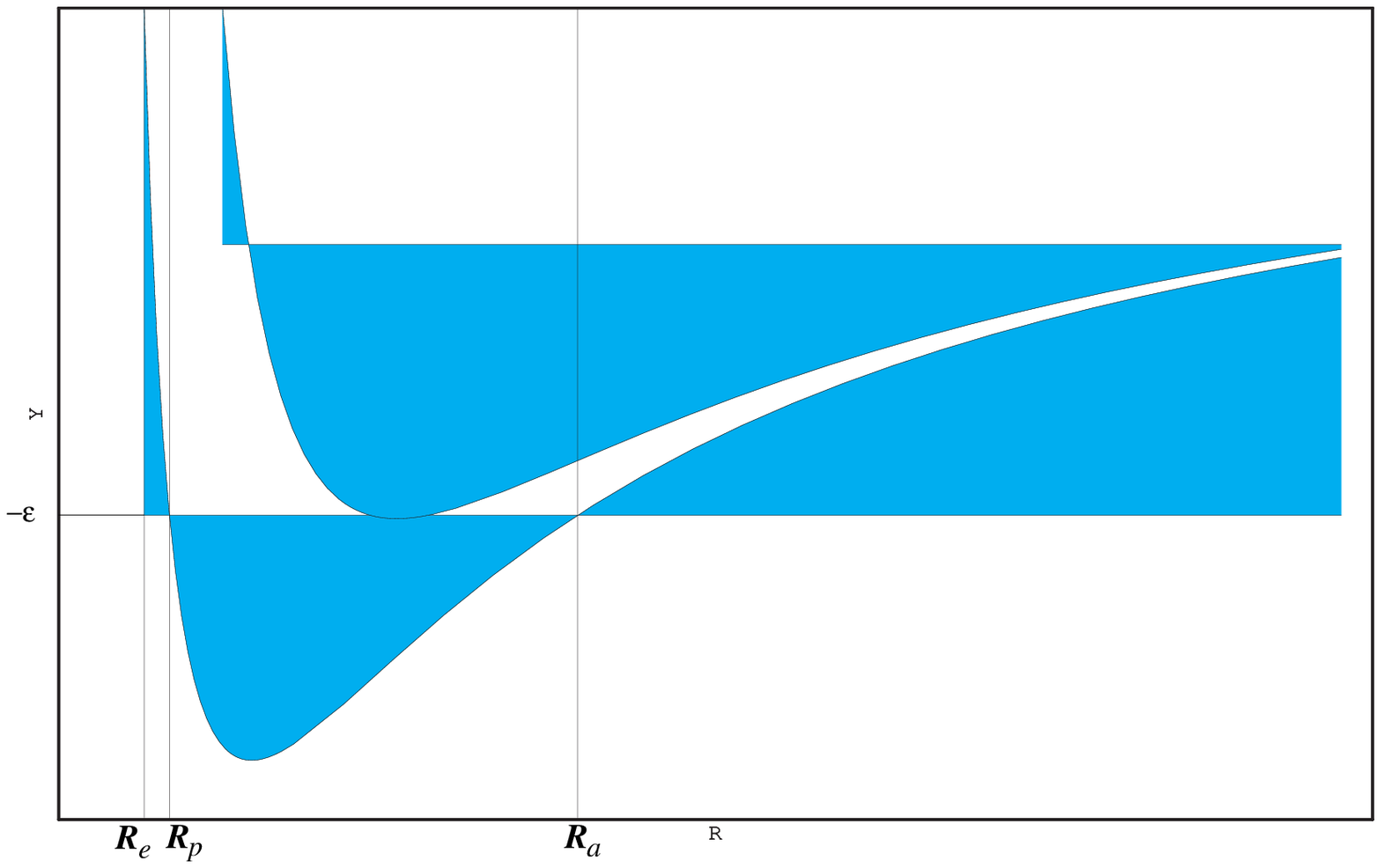}} }
\Caption
the relationship between integration in terms of the effective
 potential, $Y$, and $R$ integration.  The effective potential  is
 drawn here as a function of $R$ for two values of $h$. The area
 between that curve and the lines $Y= -\varepsilon$ and 
$R={\cal R}_e$ is shaded,
defining 3 regions from left to right. The middle region does not
contribute to \Eq{invertAbelR}.
 The equation $Y= -\varepsilon$ has two roots corresponding to the perigee
 and the apogee of the star, while $Y= 0$ has two roots corresponding
 to infinity and ${\cal R}_e$, the inner bound of an orbit with zero
 energy and angular momentum $h$. The sign of the slope of $Y(R)$
 gives the sign of the contribution for each branch in \Eq{invertAbelR}.
\endCaption
\endfigure 
\subsection{  Disks with given $Q$ profiles }
From the  point of view of stability analysis, $F_\phi$ may be chosen to match given constraints such as 
a specified  potential and  temperature profile of a disk.
The azimuthal pressure, $p_\phi=\Sigma(\langle v_\phi^2 \rangle - \langle
v_\phi \rangle^2 )$, is then fixed via the equation of radial 
support and  by the temperature of the disk defined by Toomre's $Q$ number:
$$ p_\phi= \Sigma V_c^2 +{\partial ( R p_R)\over \partial R } \, , 
\quad {\rm and } \quad p_R = Q^2 \, { \pi^2  \Sigma^3 \over \kappa^2 }\, , \EQN radialsupport$$
where the circular velocity, the epicyclic frequency and the surface density
 are given by
 $$V_c^2 = -\left[ {1\over R }{\partial  \psi\over \partial R } \right]
\, , \quad \kappa^2 ={1\over R^3 } \left[ {\partial ( R V_c)^2 \over
\partial R } \right]\,, \quad {\rm and } \quad \Sigma ={1\over 2 \pi }
\left[ {\partial \psi \over \partial z } \right] \, .\EQN temperature
$$
Note that here the azimuthal stress $p_\phi$ contains the mean
streaming stress.
It is assumed here that the field is self-consistent and obeys
\Eq{fish}; alternatively $\Sigma$ may be given independently of $\psi$
and the method described here still applies.  The surface density,
$\Sigma$, and the azimuthal pressure,
$\Sigma \langle v^2_{\phi}  \rangle $, may in turn be expressed in terms of
 $F_\phi(R,v_\phi)$ via
 $$\EQNalign{  \Sigma  & \equiv   \int
 \left[\int  {f_+(\varepsilon,h) 
 d \, v_{_R}  } 
 \right] \, d v_\phi = 2  \int \limits_0^{v_e}
 F_\phi(R,v_\phi) \, d v_\phi \, , \EQN FR;a\cr 
  \Sigma \langle v^2_{\phi}  \rangle &\equiv   \int
 \left[\int  {f_+(\varepsilon,h) 
 d \, v_{_R}  } 
 \right] \, v_\phi^2\, d v_\phi  = 2 \int \limits_0^{v_e}
 F_\phi(R,v_\phi) \, v_\phi^2 \, d v_\phi \, ,   \EQN FR;b \cr }$$
 where the circular escape velocity, $v_e$, is equal to $\sqrt{2 \psi}$.
The function $f_+$ stands for the  component of the distribution
function even  in $h$.  Any function $F_\phi$ satisfying these moment
equations  corresponds to a state of equilibrium stable
against ring formation when $Q>1$. Realistic choices for $F_\phi$ are
presented in the next sections.

\subsection{  Alternative inversion method}

Another intermediate observable, $ G_\varepsilon(R,\varepsilon)$, the
 distribution function for the number of stars which have specific
 energy $\varepsilon$, may also be parameterised to fix the surface
 density and the average energy density profiles.  For external
 galaxies $G_\varepsilon$ is not directly observed. For our Galaxy,
 $G_\varepsilon$ is measurable indirectly -- via the distribution of
 stars with given radial {\sl and} azimuthal velocity given by
 spectroscopy and proper motions -- but only yields the even component
 in $h$ of the distribution function. However, the corresponding inversion
 method still provides a route for the construction of models with
 specified temperature profiles. \nl
 The number of stars which have specific energy $\varepsilon$ within $d \varepsilon$ 
 at radius $R$ reads
$$ G_\varepsilon(R,\varepsilon) = \int {f_+(\varepsilon,h) \over v_{_R} }\, d v_\varphi
={2 \over \sqrt{2}} \,  \int \limits_{0}^{X} { g(\varepsilon,h^2/2) \over \sqrt{ X- h^2 / 2} }\,  \, 
d \,\left({h^2 \over 2}\right) \, , \EQN G$$
 where $X$ is  defined by
 $$ X= R^2( \psi + \varepsilon ) \, . \EQN defX $$
Here the auxiliary function $g$ is given by
 $$g(\varepsilon,h^2/2) ={f_+(\varepsilon,h)/ |h| }\,  . \EQN g2f $$
 The surface density, and  mean energy density,
 $\left\langle \varepsilon  \right\rangle$,
 may be expressed in terms of
 $G_\varepsilon(R,\varepsilon)$ via
 $$\EQNalign{  \Sigma  & \equiv   \int
 \int  {f_+(\varepsilon,h) 
 d \, v_\phi  }
  \, d  \, v_{_R}  =   \int \limits_{-\psi}^0
 G_\varepsilon(R,\varepsilon)\, d \varepsilon \, ,  \EQN GR;a\cr 
\Sigma \,  \left\langle \varepsilon  \right\rangle  &\equiv   \int
 \int  {f_+(\varepsilon,h) 
\left[{v_\phi^2 \over 2 }+{v_{_R}^2 \over 2 } -\psi \right] \,  d  v_{\phi}  }
 \,  \, d v_{_R}  = \int \limits_{-\psi}^0
 G_\varepsilon(R,\varepsilon) \, \varepsilon  \,  d \varepsilon  \, . \EQN GR;b \cr }$$
Note that only the symmetric 
component of the distribution function follows from \Eq{g2f}.
The local mean energy density
  $\langle \varepsilon  \rangle = 
\Sigma^{-1} ({p_R/ 2} + {p_\phi /2}) - \psi(R) $
is  fixed  by the temperature of the disk defined by Toomre's $Q$ number:
$$ \langle \varepsilon  \rangle ={1\over 2}
\left[ \Sigma V_c^2 +{\partial ( R p_R)\over \partial R } + p_R \right] - \psi(R) \quad {\rm given } \quad
 p_R = Q^2 \, { \pi^2  \Sigma^3 \over \kappa^2 }\, , \EQN energysupport$$
where $\kappa$ and $V_c$ are given by \Eq{temperature}. 
The function   $G(R,\varepsilon)$ may be expressed in terms of $\varepsilon,X$ via  
 \Eq{defX} and $R=R(\psi)$. 
 Calling  ${\tilde G_\varepsilon} (R,X) \equiv G_\varepsilon(R,\varepsilon)$ 
 leads to the inversion of  \Eq{ G} by  an Abel transform:
 $$g(\varepsilon,h^2 / 2) = {1 \over  \sqrt{2} \, \pi} \int \limits_0^{h^2/2} 
 {\partial {\tilde G}_\varepsilon \over \partial X}
   { \,d  X \over \sqrt{ h^2/2 - X} }  \, . \EQN invertGAbel $$
$f_+(\varepsilon,h)$ follows from $g(\varepsilon,h^2 / 2)$ via \Eq{g2f}.
\nl
Any function $G_\varepsilon$ satisfying the moment constraint \Eqs{GR}
yields, via \Eq{invertGAbel}  and \Ep{energysupport}, a maximally rotating disk 
stable against ring formation.

 \subsection{Implementation: Gaussian line profiles}
Gaussian velocity distributions  are desirable both as building blocks to fit measured 
line profiles and as ``realistic'' choices for $F_\phi$ in the construction scheme of disks 
parameterised by their temperature. The former point is discussed in  
section~4.
The construction of Gaussian line profiles compatible with a given temperature requires   
a supplementary assumption for the mean azimuthal velocity  of the flow, 
$\left\langle v_\phi  \right\rangle$,
on which the Gaussian should be centred. In additions to the the two
constraints,
\Eqs{FR}, this
 puts a  third constraint on $F_\phi$, namely
$$ \Sigma  \left\langle v_\phi  \right\rangle  \equiv   \int
 \left[\int  {f_-(\varepsilon,h) 
 d \, v_{_R}  } 
 \right]  \, v_\phi \, d v_\phi = 2  \int \limits_0^{v_e}
 F_\phi(R,v_\phi)  \, v_\phi \, d v_\phi \, ,  \EQN FRR$$
where $f_-$ is the odd component in $h$ of the distribution function.
Suppose the following functional form for $F_\phi$
$$ F_\phi(R,v_\phi)= S(R) \, W_n(R,v_\phi) \,
\exp\left( -{ \left[ v_\phi- v(R) \right]^2 \over 2 \sigma^2(R)} \right) \, ,\EQN formF $$ 
where the window function $W_n$ is  intended to damp $F_\phi(R,v_\phi)$ 
near the escape velocity  $v_e$. A possible expression for $W_n$ is
$$W_n(R,v_\phi) =\left\{ \eqalign{ \exp\left[ -{v_\phi^2 \over n^2
 \left( v_e^2(R)-v_\phi^2 \right)^2} \right] 
  \quad &if \quad  |v_\phi|<v_e \, ,\cr 
 0 \quad  \quad \quad &{\rm elsewhere} \, .  \cr  }
\right. 
   \EQN window $$ 
Self-consistency requires that the unknown functions $(v,\sigma,S)$ of
\Eq{formF}  are solved in terms of
$\left\langle v_\phi  \right\rangle, \, \, \Sigma,\,\,{\rm and }\,\, p_\phi $  via \Eqs{FR} 
and \Eq{FRR}.  
For practical purposes, the temperature range explored  in realistic disk models is such that 
  $  ( v_e- v )^2/ 2 \sigma^2  $ is  quite large at all radii;
if $n$ is also chosen so that at   temperature $\sigma$
$$n \gg {{v_e\,\left\langle {v_\phi } \right\rangle} \, \sigma
 \over {\left( {v_e^2-\,\left\langle {v_\phi } \right\rangle^2} 
\right)\left( {v_e+\,\left\langle {v_\phi } \right\rangle} \right)}}\, , \EQN$$
then the tail of the Gaussian function   need  not be taken explicitly into 
account. 
The line profile $F$  then reads directly in 
terms of $\left\langle v_\phi  \right\rangle,\,\Sigma, $ and $p_\phi$:
$$  F_\phi(R,v_\phi)= {\Sigma(R) \over \sqrt{2 \pi} \, \sigma_\phi } \,  
\exp\left( -{ \left[ v_\phi- \left\langle v_\phi
  \right\rangle \right]^2 \over 2 \sigma_\phi^2} \right) \, , \EQN formF2 $$
where  $\sigma_\phi$ is the the azimuthal velocity dispersion  $$\sigma_\phi^2 = p_\phi/\Sigma -
 \left\langle v_\phi  \right\rangle^2 \, . \EQN $$

The azimuthal pressure $p_\phi$ follows from the equation of radial 
support and the temperature of the disk defined by Toomre's $Q$ number given by \Eq{radialsupport}.
The expression of the average azimuthal velocity, $\langle v_\phi  \rangle$,
may be taken    to be  that 
which leads to an exact  asymmetric drift  equation:
 $$\Sigma \left\langle v_\phi  \right\rangle^2=
 p_\phi - p_R\left({\kappa^2 R^2 \over 4 V_c^2  }\right) \,. \EQN drift $$
Equations \Ep{formF2}-\Ep{drift}, 
together with \Eqs{radialsupport} provide a prescription  for the 
Gaussian azimuthal line profile $F_\phi$ which yield the distribution
function $f(\varepsilon,h)$ of a disk at given temperature profile via
\Eq{invertAbelR}.

\midfigure{DFGaussToomre}
\centerline{\hbox{\epsfysize=15cm\epsfbox{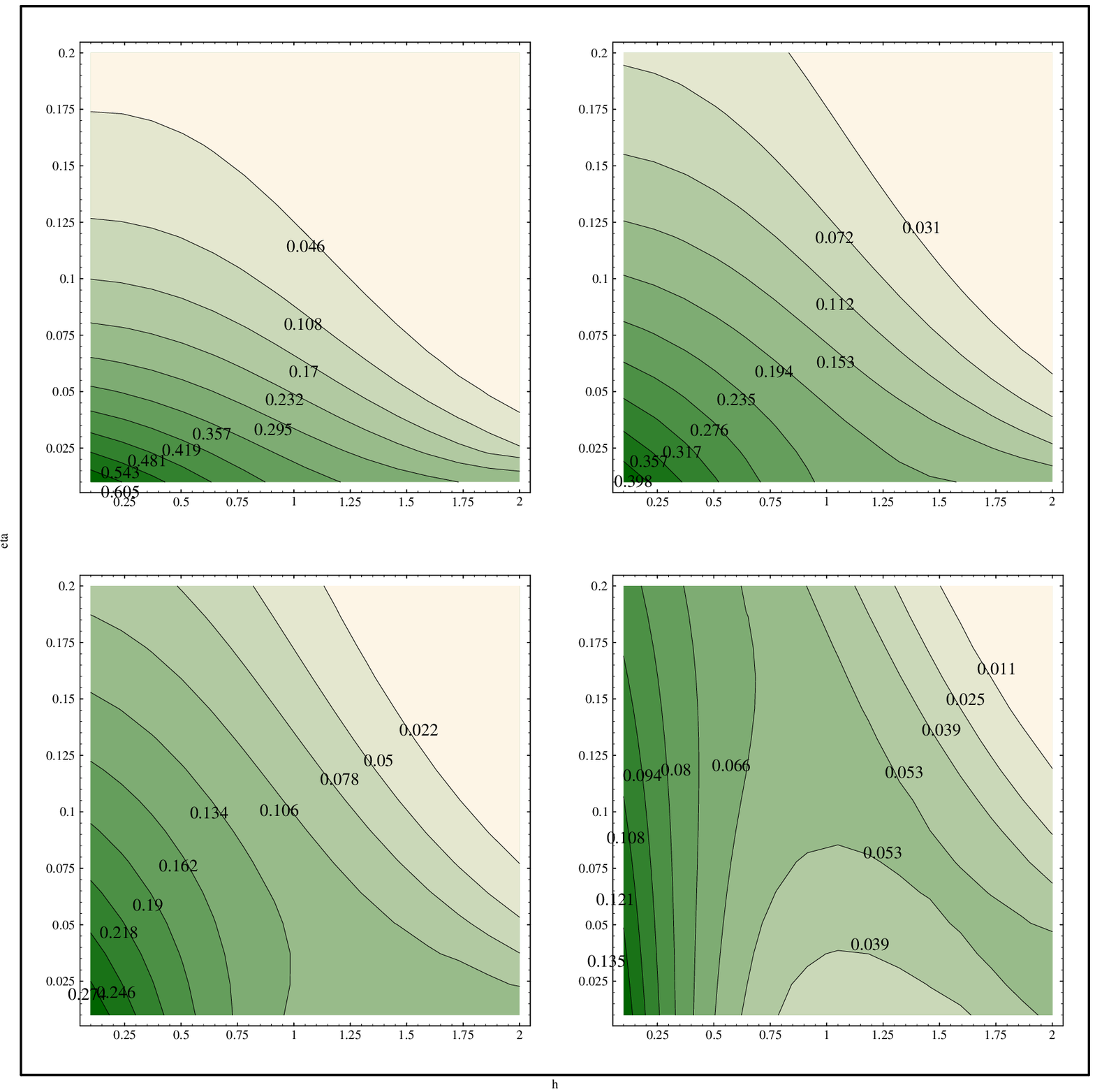}} }
\Caption
isocontours of the Gaussian distribution functions  for the Kuzmin
disk with temperature:  $Q=1$, (top left panel), $Q=1.25$, (top right  panel),
  $Q=1.75$ (bottom left panel ), $Q=2$ (bottom right panel). The
construction scheme is described by \Eq{invertAbelR} and   \Eqs{dlogFdR }.
Each diagram represents the number of star with angular momentum, $h$,
and relative energy, $\eta= \varepsilon/ \varepsilon_h$, where $ \varepsilon_h$
is the energy of the star on a circular orbit with momentum $h$.
The parametre $\eta$ therefore measures the eccentricity of the orbits.
 A hotter component is apparent at larger momentum for the $Q=2$
disk where the contours are more widely spaced.   
\endCaption
\endfigure 

 \subsection{ Example: constant temperature Gaussian Kuzmin disks}
For the Kuzmin disk, the   prescription described in 
the above for the parameters of the line profile 
yields
$$ 
\sigma_\phi= {Q \over {{{4 \left( 1 + {R^2} \right) }^{{3/  4}}}}}
\quad {\rm and } \quad 
\langle v_\phi  \rangle ={{R\,\left({{256 - 60\,{Q^2} + 128\,{R^2} - 21\,{Q^2}\,{R^2} +
 16\,{R^4}}}\right)^{1
/2} }\over 
   {4\,{{\left( 1 + {R^2} \right) }^{{3 / 4}}}\,\left( 4 + {R^2} \right) }} \, , 
$$ where $Q$ is Toomre's number.
From \Eq{formF}, $F_\phi$ becomes as a function of $h$ 
$$F_\phi[h,R]=  
   {{{\sqrt{{2/ {{{\pi }^3}}}}}}
               Q^{-1} \over  \left( 4 + {R^2} \right)}
{{\exp\left[{-{8\,{{\left( 1 + {R^2} \right) }^{{3/ 2}}} \over Q^2}\,
         {{\left( {h\over R} - {{R\,\left({{256 - 60\,{Q^2} +
128\,{R^2} -
 21\,{Q^2}\,{R^2} + 
               16\,{R^4}}}\right)^{1/2}}\over 
                {4\,{{\left( 1 + {R^2} \right) }^{{3/ 4}}}\,{\left( 4
+
 {R^2} \right) } }
                } \right) }^2}} \right]}} \, .$$
Differentiating with respect to $R$ gives
$${\partial\log F_\phi \over \partial R} = (4+R^2)\,  \left[\eqalign{&
 {{8\,{{{h}}^2}\,
      \left( 2 - {{{R}}^2} \right) }\over 
    {{Q^2}\,{{{R}}^3}\,{{\left( 1 + {{{R}}^2} \right) }^{{1/4}}}}}
+ {{4\,{R}\, P_6
      }\over 
    {{Q^2}\,{{\left( 1 + {{{R}}^2} \right) }^{{3/ 4}}}\,
      {{\left( 4 + {{{R}}^2} \right) }^3}}}  \cr &
 -{{3\,{R}}\over 
    {2\,{{\left( 1 + {{{R}}^2} \right) }^
        {{7/ 4}}}}}
   - {{6\,{h}\,  F_6 \,{R}\,
      \,
       }\over 
    {{Q^2}\,\left( 1 + {{{R}}^2} \right) \,
      {{\left( 4 + {{{R}}^2} \right) }^2}\,
     }} \cr} 
  \right] \, ,  \EQN dlogFdR $$
with 
$$\EQNalign{
 F_6&={ \left( -1024 + 216\,{Q^2} - 768\,{{{R}}^2} + 
        106\,{Q^2}\,{{{R}}^2} - 192\,{{{R}}^4} + 
        7\,{Q^2}\,{{{R}}^4} - 16\,{{{R}}^6} \right) \over \left(
{{256 - 60\,{Q^2} + 128\,{{{R}}^2} - 
          21\,{Q^2}\,{{{R}}^2} + 16\,{{{R}}^4}} }\right)^{1/2} }  \, ,
 \EQN dlogFdRP6;a \cr
 {\rm and }& \quad P_6 =-256 + 60\,{Q^2} - 192\,{R^2} + 27\,{Q^2}\,{R^2} - 48\,{R^4} - 4\,{R^6}
         \, .
 \EQN dlogFdRP6;b\cr } $$
The equations \Ep{dlogFdR} and \Ep{dlogFdRP6} together
 with \Eq{FR}  fully characterise a complete family
 of distribution functions parameterised by their temperature via
Toomre's number. \Fig{DFGaussToomre} illustrates this inversion
procedure. 
The parametrized azimuthal velocity profile is given on
\Fig{FRFphiGauss}, together with their {\sl derived } radial velocity profile.
In contrast,   \Fig{DFGaussIsochrone} corresponds to the 
same method applied to the Isochrone disk, $\psi= 1/(1+\sqrt(1+R^2))$.
Note that the relatively broader core  of the
Isochrone disk does not require as strong a hot component to match the
imposed temperature profile.

\midfigure{FRFphiGauss}
\centerline{\hbox{\epsfysize=6cm\epsfbox{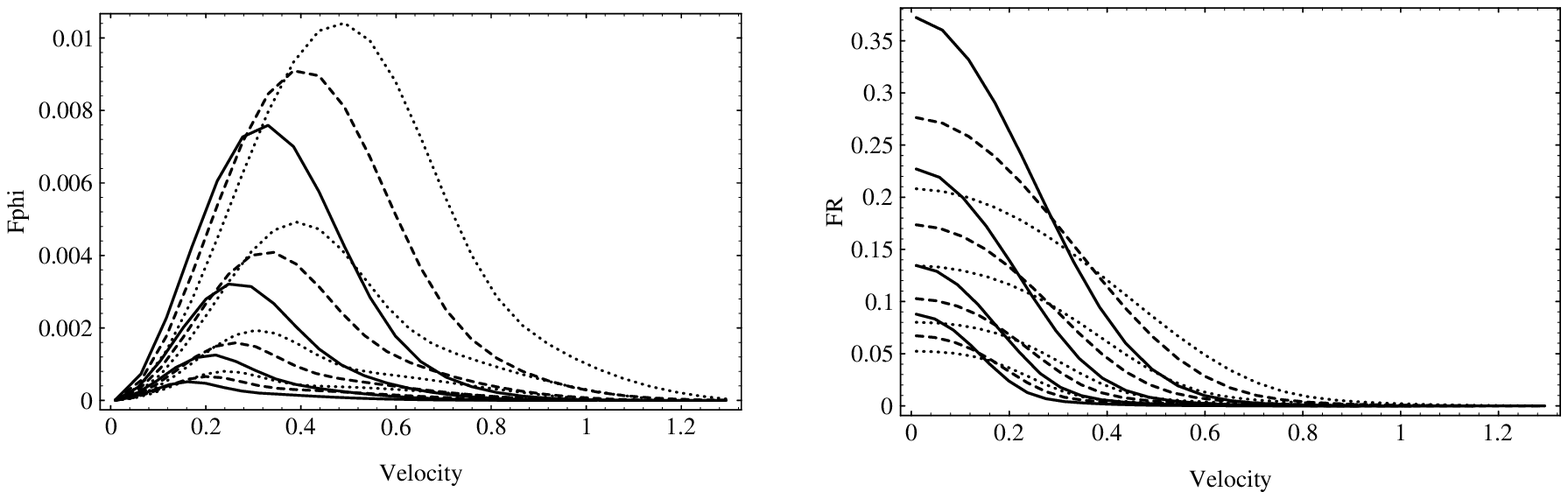}} }
\Caption
Azimuthal (left) and radial (right) velocity distributions for the Gaussian
Kuzmin-Toomre disks parametrized by their ``Q'' temperature.  The left
panel was parameterised and the right deduced 
using \Eq{FVR} and \Eq{invertAbelR}.
The curves from top to bottom correspond to radii going from
$R=0.5,1,\cdots 2.0$. The plain curve corresponds to a $Q=1$ disk, the dashed
curve to a $Q=1.25$ disk, while the  doted curves corresponds to a
$Q=1.5$ disk. Note that, as expected, the hotter disks have  broader
radial velocity distributions.
\endCaption
\endfigure 

\midfigure{DFGaussIsochrone}
\centerline{\hbox{\epsfysize=15cm\epsfbox{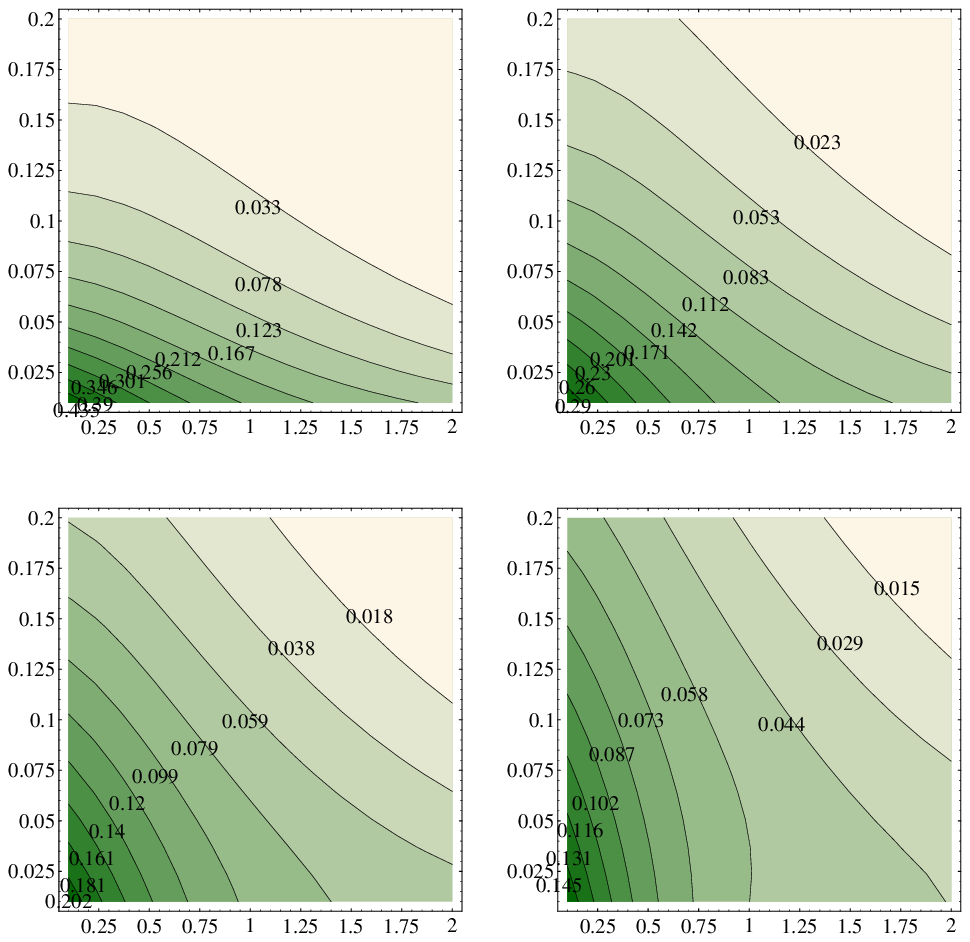}} }
\Caption
isocontours of the Gaussian distribution functions  for the Isochrone
disk displayed as in \Fig{DFGaussToomre}. 
\endCaption
\endfigure 

\subsection{ Example: Power law disks }

The potential and surface density for the power law disk read 
$$ \psi =   R^{-\beta} \quad  {\rm and } \quad  \Sigma = {S_\beta\over 2 \pi}
  R^{-\beta-1}\,,\quad \quad (0<\beta<1) \, , \EQN psiSigma$$
 where $S_\beta$ is given by :
$$S_\beta={ \beta}\, \left(\Gamma[1/2+\beta/2]\, \Gamma[1-\beta/2] \over 
 \Gamma[1/2-\beta/2]\, \Gamma[1+\beta/2]\right)\,. \EQN Sbeta $$
It is assumed 
here that distances are expressed in terms of   $R_0$, the reference radius, and
energies in terms  of  $\psi_0= \psi(R_0)$. (in units of $G=1$).
The pressures follow from \Eq{psiSigma} and the equation of radial
support \Eq{radialsupport}.
The pressures follow from \Eq{psiSigma} and the equation of radial
support
 $$p_R ={ Q^2 S_\beta^3  \over8  \pi \beta (2-\beta) }\, R^{-2\beta -1}
\, , \quad p_\phi = {\beta S_\beta  \over 2   \pi}
\left[ 1 - { Q^2 S_\beta^2  \over 2 \beta (2-\beta) }\right] \, R^{-2\beta -1}
\EQN pRpphi $$
where $Q = { \sigma_{{}_R}  \kappa /( \pi  G \Sigma})$.
Assuming that 
$ {\tilde F}_\phi (h,Y)$ scales like $  C Y^{n+1} h^b $,
the inversion yields
 $$f(\varepsilon,h) ={1 \over \sqrt{2} \pi} \int \limits_0^{-\varepsilon} 
 {\partial {\tilde F}_\phi \over \partial Y}
   { \,d  Y \over \sqrt{ (-\varepsilon) - Y} } = { (n+1) C
\Gamma(n+1) \over
\sqrt{(2 \pi )} \Gamma(n+3/2)  } (-\varepsilon)^{n+1/2} \, h^b \, , \EQN invertAbel $$
The surface density, $\Sigma$,
and  the azimuthal pressure, 
$\Sigma \langle v^2_{\phi}  \rangle $, read in terms of
 ${\tilde F}_\phi(h,Y)$ via
 $$\EQNalign{  \Sigma  &    = 2  \int \limits_0^{v_e}
 F_\phi \, d v_\phi =  {   C \,  2^{b/2+1/2}\,
\Gamma(2+n)\Gamma(b/2 + 1/2 ) \over (n+1)^{-1} \,
\sqrt{(2 \pi )} \Gamma(b/2+ n+5/2)  }\, R^{-\beta (3/2+n+b/2)-b}  \, , 
\EQN FRpl;a\cr 
  \Sigma \langle v^2_{\phi}  \rangle &= 2 \int \limits_0^{v_e}
 F_\phi \, v_\phi^2 \, d v_\phi  =  {  C \,  2^{b/2+5/2}\,
\Gamma(2+n)\Gamma(b/2 + 3/2 ) \over  (n+1)^{-1} \,
\sqrt{(2 \pi )} \Gamma(b/2+ n+7/2)  }\, R^{-\beta (5/2+n+b/2)-b-1} \,
,   \EQN FRpl;b \cr }$$
where  the circular escape velocity, $v_e$, is equal to $\sqrt{2 \psi}$.
Identifying \Eqs{FRpl} with \Eqs{psiSigma}-\Ep{pRpphi} yields the relations
$$ b = {2 \beta n \over 2 -\beta} -1\, , \quad n = 
 \beta -2 + {{2\,{{\left(  \beta-2 \right) }^2}\,
       \beta}\over {{{{\it Q}}^2}\,{S_\beta^2}}}\, , \quad
 C = { S_\beta 2^{-\beta/2-1/2}\Gamma(n+\beta/2 + 5/2 )
 \over 2 \pi \Gamma( 2+n) \Gamma(\beta/2+1/2) }\, , \EQN nbeq $$
which gives  the distribution function 
$$f_+(\varepsilon,h)=S_\beta\,  \left({-\varepsilon}\right)^{n+1/2} \, |h| \,  \left[
 {  { 2^{-5/2}\,\pi^{-3/2}  } \, \Gamma[\alpha+1] \over
  \Gamma[n+3/2] \Gamma[\alpha -n-1]}\right] \left({h^2\over
2}\right)^{\beta n /(2-\beta) -1}
\, \EQN  DFpowerlaw$$
where  $\alpha= { 1 +2 n /( 2-\beta)}  $.
 Here, the effect of the parameter $n$ on temperature becomes  clear
 when inverting \Eq{nbeq} for Q:
$$Q = {2-\beta\over S_\beta }\left[ 2\beta \over{  2-\beta+n} \right]^{1/2}\, .\EQN $$
These results for the power-law disks  were also  derived
by Evans\cite{Evans} who used the Ansatz II described in section~3.1
and illustrated in section 3.3.

\section{ Distribution functions for a given surface density}

As mentioned in section~1, the inversion problem for flattened disks
has no unique solutions.  Here the freedom in choosing the
distribution function is exploited to choose special {\it forms} of
distribution function which make the solution of the resulting
integral equation remarkably simple.  Explicit inversions are given
for distribution functions whose even parts are of the form
$f_+(\varepsilon, h) = (-\varepsilon)^{n+{1\over2}} G_n ({h^2 /2})$
where $h$ is the specific angular momentum and $\varepsilon$ the
specific energy of a given star.  Inversions are also provided when
$f_+ (\varepsilon, h) = h^{2n} F_n (\varepsilon)$.  Examples are given
and illustrated graphically .  The part of $f$ which is antisymmetric
in $h = Rv_\phi$ cannot be determined from the surface density alone
unless it is assumed that all the stars rotate in the same sense about
the galaxy, in which case $f_- = {\rm sign} (h) f_+$.  More generally,
if the mean velocity $\left\langle {v_\phi } \right\rangle$ is
specified -- say by the asymmetric drift prescription -- a similar
Ansatz for $f_-$ allows its explicit determination too.  The inversion
formulae require expressing powers of $R$ times the surface density
$\Sigma (R)$ as functions of either the potential $\psi (R)$ or of the
related function $Z(R) = R^2\psi$.  These formulae can only be applied
to those density distributions for which the potential in the plane of
the matter is known (either self-consistently or not).  The other
characteristics of the disk follow from the knowledge of $f$ and
cannot be specified {\`a}-priori.
\subsection{ Derivation}
\item{$\bullet$} {\sl  Even component of $f$: first Ansatz}

A first Ansatz is to look for solutions for $f_+$ which take the form

$$I \quad \quad \quad f_+ (\varepsilon, h) = (-\varepsilon)^{n+{1\over2}} |h| G_n  \left({h^2\over2}\right)
\, , \EQN 
2.5$$  where $n$ is first assumed to be an integer.{
Here the $|h|$ is taken out for convenience; its appearance in no way 
implies that $f_+(\varepsilon, 0) = 0$ because $G_n ({h^2 / 2})$ may behave 
as $|h|^{-1}$ for small $h$.}  The quantity $n$ measures the level of 
anisotropy in the disk. Large $n$ correspond to cold, centrifugally supported disks;
small $n$ correspond to isotropic, pressure supported disks.
 Inserting the Ansatz into \Eq{2.2} gives 
$$\Sigma (R) = {2^{3\over2}\over R} \int^{R^2\psi}_0 G_n \left({h^2\over2}\right) \left[ \int^Y_0 
{(-\varepsilon)^{n+{1\over2}}d(-\varepsilon)\over\sqrt{Y-(-\varepsilon)}} \right]
d\left({h^2\over2}\right) \, .\EQN 2.6$$ 
The inner integral can be transformed to $Y^{n+1} I_n$ where
$$I_n = \int^1_0 {x^{n+{1\over2}} dx\over\sqrt{1-x}} = { \sqrt{\pi}\, \Gamma( n+{3 \over 2} ) \over 2 \Gamma(n + 3) } = {(2n+1) 
(2n-1)...1\over2^{n+2}(n+1)!} \, \pi \, , \EQN 2.8$$
with  $x =(-\varepsilon/Y)$.
 Re-expressing $Y^{n+1}$  in terms of the potential,
writing $H$ for $h^2/2$ and $Z$ for $R^2\psi$, this expression  becomes on 
multiplication by $R^{2n+3}$
$$S_n \equiv R^{2n+3} \Sigma(R) = 2^{3\over 2} I_n \int^Z_0 (Z-H)^{n+1} G_n(H) \, 
dH \, , \EQN 2.10$$ hence defining $S_n$.
$Z\equiv R^2\psi$ is known, so $R^{2n+3} \Sigma(R)$ can be re-expressed 
as a function $S_n(Z)$.  Differentiating \Eq{2.10} $n+2$ times with respect 
to $Z$,using \Eq{2.8} and substituting $H$ for $Z$ gives:
 $$G_n (H) = {1\over\sqrt 2[(n+{1\over2})(n-{1\over2})....{1\over2}]\pi} 
\left({d\over dH}\right)^{n+2} S_n(H)\, .\EQN 2.12$$
\Eq{2.12} specifies the part of the distribution function which is  even in $h$. 
 Demanding 
{\it no} counter-rotating stars  would imply, for instance, $f = f_+ + f_- = 0$ for 
$h < 0$.  In that case $f_- = -f_+$ for $h < 0$ which leads to $f_- = 
{\rm sign}(h) \,  f_+ (\varepsilon, h)$.
 These solutions are called maximally rotating disks.
The formal solution, \Eq{2.12}, must be  non-negative for 
all $h$ so that \Eq{2.2} corresponds to a realizable distribution of stars.  
Within that constraint, the choice of $n$ is, in principle, free.  
There are many distribution functions which give the same surface density distribution 
because there are two functional freedoms in $f(\varepsilon, h)$ and only one function 
$\Sigma (R)$ is given.  Here it has been shown that even within the special 
functional form of  Ansatz I there is potentially a solution for any chosen integer $n$.  
When $n$ is not an integer, $ n = n_0 - \alpha,\, \,  0< \alpha < 1 $,
\Eq{2.6} may still be inverted using Abel transforms, though 
the final solution involves an integral which might require
 numerical evaluation
$$ G_n(H) = {\sin( \pi \alpha)  2^{-3/2} \pi^{-1} I_n^{-1} 
\over [(n_0+1-\alpha)(n_0-\alpha)...(1-\alpha)]   }
\left[\eqalign{
 &\int_0^H \left({ d  \over  d Z }^{n_0+2} S_n \right)
{ d Z \over (H-Z)^{1-\alpha}} \cr  &+ 
\left({ d  \over  d Z }^{n_0+1} S_n \right)_{Z=0}
{ 1 \over H^{1-\alpha}}  \cr }
\right] \, , \EQN contna $$
where 
$ I_n $ is defined for non integer $n$ by the first identity in \Eq{3.4}. 
  Alternatively, there are many ways in which it is 
possible to re-express $\Sigma (R)$ in the general form
$\Sigma (R) = \sum_n R^{-2n-3} S^*_n (R^2\psi)$.
For any particular sum, a distribution 
function $G^*_n (H)$ can be found which gives rise to that part of the density given by 
the n$^{\rm th}$ term $S^*_n$.  Indeed the relationship between $G^*_n$ and 
$S^*_n$ is just that given by \Eq{2.12} between $G_n$ and $S_n$.  Since the 
original integral equation is linear, it follows that each expression for 
$\Sigma (R)$  yields a distribution function in the form
$f_+ (\varepsilon, h) = \sum_n (-\varepsilon)^{n+{1\over2}} |h| G_n^* 
\left({h^2/2}\right)$.
There is no requirement that each $G^*_n$ should be positive provided that 
the sum $f_+$ is positive for all $\varepsilon$ and $h$.  Thus
 these expression for the density
and distribution functions  gives a very considerable 
extension of the original Ansatz \Eq{2.5}.

\item{$\bullet$} {\sl Odd Component of $f$ and asymmetric drift}

When the average azimuthal velocity field is known, a similar inversion 
procedure is available in order to specify the odd part in $h$ of the distribution function.
This field may be assumed to be given by the asymmetric
drift equation which takes the form $$\Sigma \left\langle v_\phi  \right\rangle^2=
 p_\phi - p_R\left({\kappa^2 R^2 \over 4 V_c^2  }\right) \, , \EQN drift $$
where the azimuthal pressure, $p_\phi$,  and the radial
 pressure, $p_R$, are derived  from the even component of $f_-$, 
and the epicyclic frequency and the circular velocity $\kappa$ and $V_c$ follow 
from $\psi$. 
Formally, the only difference from  the previous 
analysis is that in place of \Eq{1.3}, the integral 
equation relating $f_-$ and $\left\langle {v_\phi } \right\rangle$ becomes
$$R \Sigma\, \left\langle {v_\phi } \right\rangle = 4 \int^{\sqrt{2R^2\psi}}_0 \int^0_{{h^2\over 2R^2}-\psi} 
{f_-(\varepsilon, h)\over\sqrt{2(\varepsilon +\psi)R^2 -h^2}} h\, d\varepsilon\, dh \, . \EQN 
4.2$$
Replacing Ansatz I by Ansatz I$'$:
$$I' \quad \quad \quad f_- (\varepsilon, h) = (-\varepsilon)^{n+{1\over2}} {\widetilde G}_n 
\left({h^2\over2}\right) \, {\rm sign} (h) \, , \EQN 4.3$$
yields a solution for ${\widetilde G}_n$ which is formally
equivalent to  \Eq{2.12}, but 
with ${\widetilde S}_n$ defined by
${\widetilde S}_n (Z) = R^{2n+4} \Sigma  \, \left\langle {v_\phi } \right\rangle$.
The extension to solutions for $f_-$ when 
$\Sigma \, \left\langle {v_\phi } \right\rangle$ is a linear
combination of  $
 R^{-2n-4} {\widetilde S}_n^* (Z) $ is straightforward.
\item{$\bullet$}{\sl Alternative Ansatz } \nl
In place of Ansatz I, consider a different Ansatz:
$$II \quad \quad \quad f_+ (\varepsilon, h) = h^{2n} F_n (\varepsilon) \, , \EQN 3.1$$
 where $n$ is also first assumed to be an integer.
Inserting \Eq{3.1} into \Eq{2.2}, and reversing the 
order of the integrations, yields
$$\Sigma (R) = 4 \int^0_{-\psi} F_n(\varepsilon)
\left[ \int^{X^{1/2}}_0 {h^{2n} 
dh\over\sqrt{X-h^2}} \right] d\varepsilon \, , \EQN 3.2$$ where
$X$ was defined in \Eq{defX}.
The inner integral is $X^n \, J_n$, where
$$J_n = \int^1_0 {x^{2n} dx\over\sqrt{1-x^2}} =  
{{{\sqrt{\pi }}\,{\Gamma}({1\over 2} + n)}\over {2\,{\Gamma}(1 + n)}}=
[(n-{\textstyle{1\over2}})(n-{\textstyle{3\over2}})...{\textstyle{1\over2}}] \pi/(2 n!) \, ,
 \EQN 3.4$$ with
$x = h/X^{1\over2}$. 
Thus \Eq{3.2} becomes, using \Eq{defX} for $X^n$
$$\Sigma (R) = 2^{n+2} J_n R^{2n} \int^0_{-\psi} (\varepsilon + \psi)^n F_n 
(\varepsilon) \, d\varepsilon \, . \EQN 3.5$$
Re-expressing $R^{-2n} \Sigma(R) \equiv {\cal S}_n (\psi)$ and writing
$\varepsilon$ for $-\psi$ yields 
$$F_n (\varepsilon) = 
{2^{-n-1}\over[(n-{1\over2})(n-{3\over2})...{1\over2}]\pi} \left({d\over 
d(-\varepsilon)}\right)^{n+1} \left[{\cal S}_n(-\varepsilon)\right]\, ; \EQN 3.7$$
 Equation \Eq{3.7} was derived independently by
Sawamura (1987)\cite{Sawamura}.
\nl
When $n $ is not an integer, the solution to \Eq{3.5} can  still be found, 
though with a supplementary integration (with $ n = n_0 - \alpha$, $n_0$ the
closest upper integer), to give
$$ F_n(-\varepsilon) ={  2^{-n -2} \pi^{-1} J_n^{-1}
  \sin \pi \alpha \over  [(n_0+1-\alpha)(n_0-\alpha)...(1-\alpha)]
} 
\left[\eqalign{ &
{\int^\varepsilon_0 \left({d ^{n_0+1}   \over d \psi  }{{\cal S}_n}\right)  
 {d \psi \over [\varepsilon - \psi]^{1-\alpha} }} 
\cr &
+{1\over \varepsilon^{1-\alpha} }
\left({d ^{n_0}   \over d\psi  }{{\cal S}_n} \right)_{\psi =0} \cr 
}\right]\, ,\EQN $$ where 
$ J_n $ is defined for non integer $n$ by the first identity in \Eq{3.4}. 
The Ansatz in  \Eq{3.1} may be also  generalised when $\Sigma (R)$ is expressed 
as linear combinations of $ R^{2n} {\cal S}^*_n (\psi)$.
\item{$\bullet$}{\sl Alternative Ansatz: odd component } \nl
The above procedure is also straightforward to implement in 
order to constrain the antisymmetric component of the 
distribution function. Indeed   Ansatz II$'$:
$$II' \quad f_- (\varepsilon, h) = h^{2n-1} {\widetilde F}_n 
(\varepsilon)\, , \EQN 4.7$$ may be chosen in  place of Ansatz I$'$.
Then  \Eq{3.2} becomes
$$R\, \Sigma\, \left\langle {v_\phi } \right\rangle = 4 \int^0_{-\psi} {\widetilde F}_n (\varepsilon) \int^{X^{1/2}}_0 
{h^{2n}dh\over\sqrt{X-h^2}} d\varepsilon \, . \EQN 4.8$$
Hence the solution for $F_n$ is formally identical to  \Eq{3.7} but with ${\widetilde {\cal S}}_n$ 
substituted for ${ {\cal S}}_n$, where
${\widetilde {\cal S}}_n(\psi) = R^{1-2n} \Sigma  \, \left\langle {v_\phi }
 \right\rangle$.
Again, the integral equation is linear, so the prescription defined by \Eq{3.7}
 when $\Sigma \, \left\langle {v_\phi } \right\rangle$ can expressed as
 linear combinations of $ R^{2n-1} {\widetilde {\cal S}}_n^* (\psi)$.

\subsection{Disk properties}
\item{$\bullet$} {\sl Radial velocity dispersion}

The average radial velocity dispersion, $\sigma_R^2$, defined by 
\Eq{2.2-2} is an important 
quantity for the local stability of the disk. 
Given \Eq{2.5} and using Ansatz I,
  \Eq{2.2-2} may be rearranged as
$$ \Sigma \sigma_R^2 ={ 2^{5\over2} {\cal I}_n\over R^{2n+5}} \int^{Z}_0 
({Z}-{H})^{n+2} G_n({H}) \, 
d H \, . \EQN 5.3$$
The calculation of $\sigma_R^2(R)$ 
is therefore straightforward once the function $G_n$ has been found.
Note the similarity between \Eq{5.3} and  \Eq{2.10}. In fact this similarity provides a check 
for self-consistency since one should have 
$${\partial\over  \partial Z}\left( 
\sigma_R^2 \Sigma R^{2n+5} 
 \right)= 2(n+2){{\cal I}_n / I_n} R^{2n+3} \Sigma=   R^{2n+3} \Sigma \, . \EQN $$
Putting  \Eq{3.1} into \Eq{2.2-2} yields, for Ansatz II:
$$ \Sigma\, \sigma_{R}^2 = 2^{n+3} {\cal J}_n R^{2n} \int^0_{-\psi} (\varepsilon + \psi)^{n+1} F_n 
(\varepsilon) \, d\varepsilon \, . \EQN 33.5$$
The similarity between \Eqs{3.5} and \Ep{33.5} yields  the identity:
$$ {\partial \over  \partial \psi } \left(\Sigma \sigma_R^2 R^{-2n}  \right)
 = 2 (n+1) \Sigma R^{-2n} { {\cal J}_n / J_n } = \Sigma R^{-2n} \,.
\EQN ident2 $$
\item{$\bullet$}{\sl Azimuthal velocity dispersion }  

The average azimuthal velocity dispersion, $\sigma_\phi$, is an observable constraint on 
models which satisfies
$$ \EQNalign {\Sigma  \left(\sigma_\phi^2 + \left\langle {v_\phi } \right\rangle^2 \right) =
\Sigma \, \left\langle {v_\phi^2 } \right\rangle &=
 { 4 \over R^3 }  \int \limits_{0}^{\sqrt{2 R^2 \psi}} \int \limits_{{ h^2 \over 2 R^{2} } 
-\psi }^{0}
{{ f_+ 
(\varepsilon, h) \, h^2 \, d\varepsilon \, dh \over
 \sqrt{2(\varepsilon +\psi)-{h^2\over R^2} }}}\, ,  \EQN 7.1 \cr}$$
where the mean azimuthal velocity $\left\langle {v_\phi }
\right\rangle$ is
 given in terms of $f_-$ by 
 linear combinations of \Eqs{4.2} and  \Ep{4.8}.
Putting \Eq{2.5}  into 
  \Eq{7.1} and following the substitutions \Eq{2.5}-\Ep{2.12}
yields, for the first Ansatz,
$$ R^{2n +5}\, \Sigma  \left(\sigma_\phi^2 + \left\langle {v_\phi } \right\rangle^2 \right) = 2^{5 \over 2} I_n \int^Z_0 (Z-H)^{n+1}  H G_n(H) \, 
dH \, . \EQN sigmaphi1$$ \nl
Putting  \Eq{3.1} into \Eq{7.1} gives, for the second Ansatz:
$$\Sigma \, \left\langle {v_\phi^2 } \right\rangle = 2^{n+4} J_{n+1} R^{2n} \int^0_{-\psi} 
(\varepsilon + \psi)^{n+1} F_n 
(\varepsilon) \, d\varepsilon \, . \EQN 3.5$$ 
Note that $\sigma_{R},\,\,\psi,\,\,\Sigma \,\, {\rm and}\,\,\langle {v_\phi^2 } \rangle$
are related via the equation of radial support \Eq{radialsupport}.
The properties of the disks constructed with Ansatz I \& II are
illustrated in \Figs{kuzminDF} and \Fig{ToomreKuzmin} for the examples
described below.
\subsection{Examples }
\item{$\bullet$}{\sl Isochrone disks } 

 Distribution 
functions for the flat Isochrone
are simplest with the assumptions of Ansatz II.
In the equatorial plane, the Isochrone potential reads 
$$ \psi =GM/(b+{r_b}) \quad {\rm where } \quad {r_b}^2 = R^2 + b^2  \, . \EQN 6.1$$
The corresponding surface density is
$$\Sigma ={{Mb} \over {2\pi R^3}}\left\{ {\ln [(R+{r_b})/b]-R/{r_b}} \right\}={{Mb} 
\over {2\pi R^3}}L  \, .
\EQN 6.2$$
But $R^2\psi =Z = ({r_b}^2-b^2) GM/({r_b}+b) =GMb(s-1)$,
where $s$ is the dimensionless variable ${r_b}/b$. Therefore
$$ \Sigma R^{2n+3} = {Mb \over 2 \pi} R^{2n} L(s)= {Mb^{2n+1} \over 2 \pi}
(s^2-1)^n L(s) \, , \EQN 6.3$$
where $ L(s) = \ln(\sqrt{s^2-1}+s) +\sqrt{s^2-1}/s$. {
Note that $L'(s)$ is quite simple: $L'(s) = \sqrt{s^2-1}/s^2$}.
With this notation,
$$\left( {{\partial  \over {\partial Z}}} \right)^{n+2}(R^{2n+3}\Sigma )=
{{Mb^{2n+1}} \over {2\pi (Gmb)^{n+2}}}\left( {{\partial  \over {\partial s}}} 
\right)^{n+2}\left[ {(s^2-1)^nL(s)} \right] \, , \EQN 6.4$$
which implies, in turn, that 
$$G_n({h \over 2}^2)={{M^{-n-1}b^{n-1}} \over {2^{5/2}\pi GI_n(n+1)\! }}\left. {\left( {{{\partial ^{}} \over {\partial ^{}s}}} \right)^{n+2}[(s^2-1)L(s)]} \right|_{s = 1+h^2/2}
\, . \EQN $$
Therefore, the final distribution function (or rather its symmetric part, $f_+$)  reads
$$f_+(\varepsilon,h)={\left[ {{{-\varepsilon } / 
{(GM/b)}}} \right]^{n+1/2} \over {4\pi ^2Gb(n+1/2)!! }}\left( {{{h^2} \over {2GMb}}} \right)^{1/2}\left(
 {{{\partial ^{}} \over {\partial ^{}s}}} \right)^{n+2}[(s^2-1)^nL(s)]\, .  \EQN 6.5$$
with $s =1 +h^2/(2GMb)$.
Using these distributions functions, Toomre's Q 
criterion for radial stability ($ Q =  \sigma_{{}_R}  \kappa /  3.36
G \Sigma_0   $)  may  be calculated via \Eq{5.3}.
 Note that for the Isochrone potential
$$ \sigma_R^2(R) ={2^{5/2} {\cal I}_n \over R^{2n+5} \Sigma(R) }
\int_0^{Z} (Z-H)^{n+2} \,
\left[
\left( \partial \over \partial s\right)^{n+2} L(s) (s^2-1)^n
 \right]_{s=H+1}
\, d\,H \, . \EQN 6.6$$
\midfigure{ToomreKuzmin}
\centerline{\hbox{\epsfysize=10cm\epsfbox{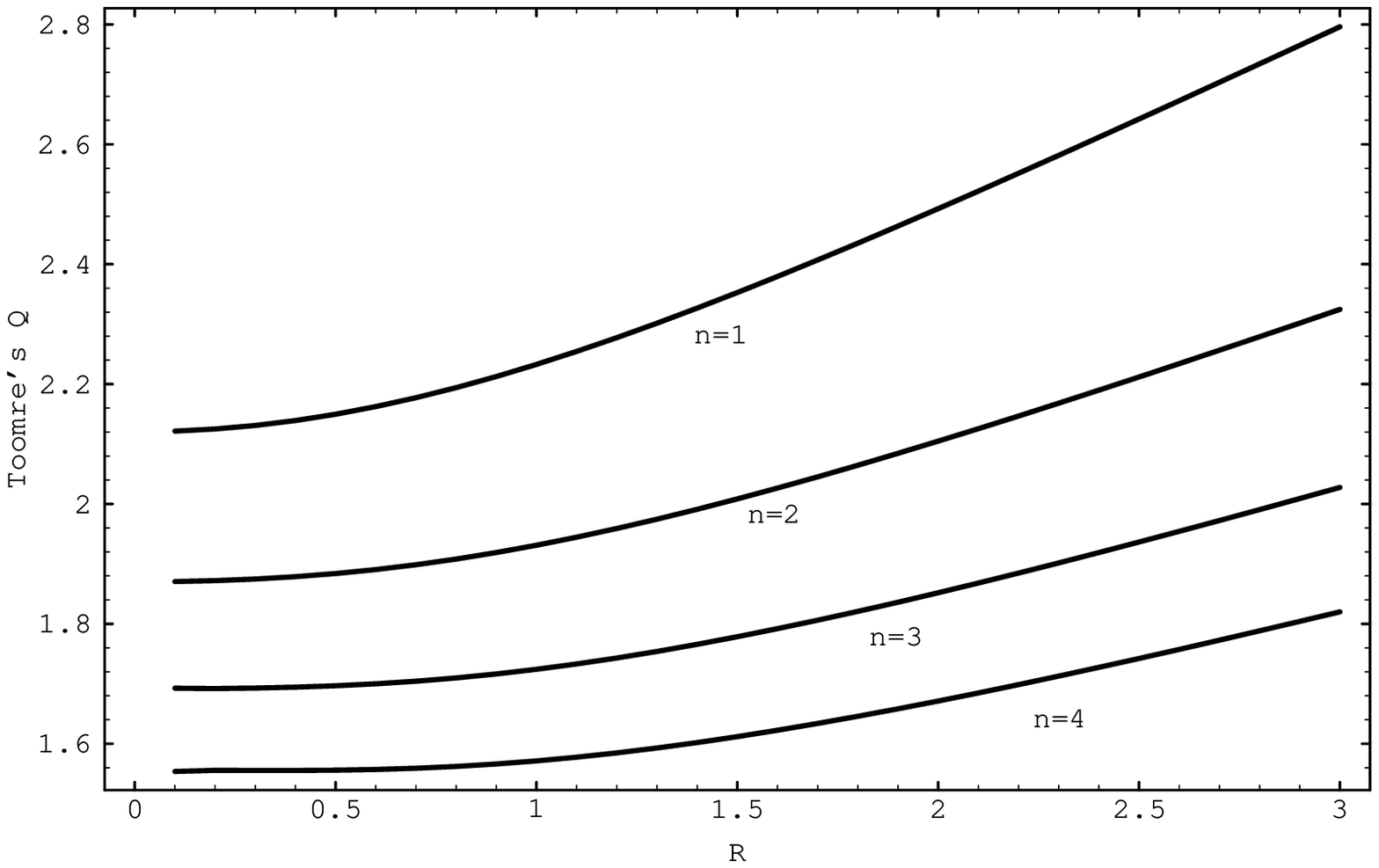}}}
\Caption
  the "Toomre Q number", $Q=  \sigma_{_R}\kappa / (3.36 \Sigma_0 )  $, 
 against radius $R$ for the Isochrone model
described by \Eqs{6.5} and \Ep{6.6}. The parameter $n$ in \Eq{6.5}
corresponds to a measure of the temperature of the disk.
\endCaption
\endfigure 
\item{$\bullet$}{\sl Toomre-Kuzmin disks}

Ansatz II is more appropriate when seeking 
distribution functions for the Toomre-Kuzmin disks.
In the equatorial plane, the potential of the disk reads 
$ \psi = -{GM/ r_b } $,
while the surface density is
$ \Sigma = {( 2 \pi)^{-1} }{GM \, b  / r_b^3}$.
Ansatz II yields the distribution (in units of $b$ and $GM/b$ )
$$ f^n_+(\varepsilon,h)={2^{n-2} \, h^{2n} \over [(n-{1\over2})(n-{3\over2})...{1\over2}]\pi^2 }
\left[{s^{2n+3}\over (1-s^2)^n}\right]^{(n+1)}_{s\rightarrow
-\varepsilon} \, . \EQN 3.1 $$
This example illustrates the drawbacks of Ansatz II. 
 The factor $h^{2n}$ in \Eq{3.1} removes all zero angular momentum orbits. 
Self-consistency yields a solution 
with circular orbits near the angular momentum origin ($f$ scales like powers of 
$h^2/ [1- (-\varepsilon)^2]^{3/2} $). In position space, this implies
that the inner core of the galaxy is made of circular orbits!
Realistic distribution functions will therefore require superpositions
of solutions of the family defined by \Ep{3.1},  as illustrated by \Fig{kuzminDF}. 
It can be shown from  \Eqs{3.1} and \Eq{3.5} that $\Sigma \, \sigma^2_{R} $
does not fall off fast enough to yield cold disks (it ought to scale
like $\psi^6$ to lead to constant $Q$ profiles).
\midfigure{kuzminDF}
\centerline{\hbox{\epsfysize=10cm\epsfbox{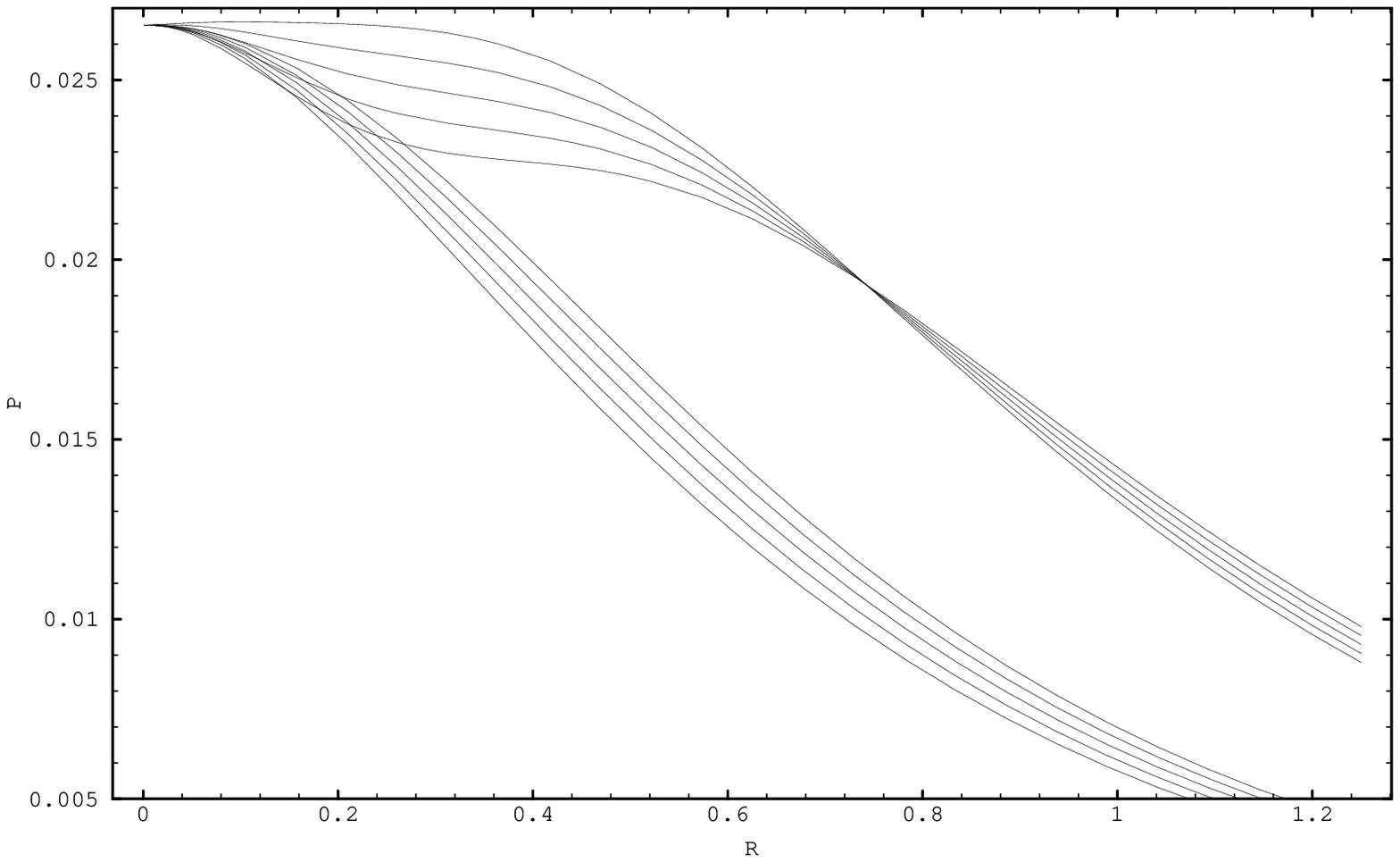}} }
\Caption
radial (bottom lines) and azimuthal (top lines) pressure for a Kuzmin
disk constructed by linear superposition of \Eqs{3.1}. 
The weights are $(2/3, 1/12, 1/4 - i/12, i/24, i/24)$ where $i=10, 
13 \cdots 22 $ for the $n=0,1\cdots$ models.
It was checked
that the superposition has positive mass everywhere. 
\endCaption
\endfigure 
\item{$\bullet$}{\sl Power law disks  }

The simplicity of power law disks leads to solutions for the inversion
problem which may be extended to models with a continuous free
``temperature'' parameter as described above.   This
temperature may  be varied continuously between the two
extremes corresponding to the isotropic and cold disk.
The potential and surface density for the power law disk where given
by \Eqs{psiSigma}--\Ep{Sbeta}.
The  inversion method  corresponding to Ansatz~I may be carried out 
and, using  \Ep{ contna}, yields  the distribution function found in \Eq{DFpowerlaw}.
The velocity dispersions are recovered from \Eqs{5.3} and \Ep{7.1}:
$$ \sigma_\phi^2 + \left\langle {v_\phi } \right\rangle^2 ={n \, \beta
\over ( 2-\beta+n) } { R^{-\beta} }\, , \quad \sigma_R^2
={2-\beta \over 2 \, ( 2-\beta+n) }
\, 
{ R^{-\beta}}\, . \EQN disp $$
  By construction, these dispersions satisfy  the equation of  radial equilibrium:
$$ \left\langle v_\phi^2  \right\rangle-  R {\partial \psi \over \partial R} = 
-{\beta \,( 2-\beta ) \over  ( 2-\beta+n) }
\, 
{ R^{-\beta}}=
 {\partial \left( R \Sigma \sigma_R^2 \right) \over \Sigma \,  \partial  R}\, .
 \EQN  $$

 \section{ Observational implications }
The inversion method described in  section~2 may be applied on observational
data which are  now   becoming available ({\it  e.g.}  Bender  {\it  et al.}
(1994)\cite{bender} and  Fisher (1994)\cite{fisher}  have presented line  of
sight  velocity distributions of elliptical   and S0 galaxies obtained  from
spectra with  (60   km/s)  resolution  and ${\rm  S/N  \sim  50}$   ).   The
observations  may be  carried  out as follows:   consider a disk galaxy seen
almost  edge   on.  It should  be chosen   so  that it   looks approximately
axisymmetric and flat. It  may contain a fraction of  gas in order to derive
the   mean   potential $\psi$  from    the  observed  H  I    velocity curve
(Alternatively, the potential may be derived directly from the kinematics if
the  asymmetric drift assumption  holds).  Putting a  slit on its major axis
and  cross  correlating the derived  absorption  lines with template stellar
spectra yields  an estimate  of  the projected  velocity distribution  which
should   essentially    correspond  to   the      azimuthal  line    profile
$F_\phi(R,v_\phi)$. An estimate   of  the radial line profile   $F_R(R,v_R)$
arises when the slit is placed along the  minor axis.  But $F_R(R,v_R)$,
is given by \Eq{FVR} while \Eq{invertAbelR} gave $f$ expressed as a function
of $F_\phi(v_\phi,R)$.    Therefore, putting \Eq{invertAbelR}  into \Eq{FVR}
provides a  simple way  of  predicting  the  radial   line profiles if   the
azimuthal line profiles are given. This procedure is illustrated on
\Fig{FRFphiGauss} for simulated data without noise.
  The    surface  density  $\Sigma$ follows    from   both $F_R(v_R,R)$  and
$F_\phi(v_\phi,R)$ and could  also be compared  with the  photometry of that
galaxy.  The likely discrepancy between the predicted  and the residual data
may be used to asses the limitations  of the reconstruction scheme.  Indeed,
the above prescription for  the inversion relies on a  set of hypotheses for
the nature of the flow (axial symmetry, thin disk approximation {\it etc..})
and   the quality of measurements.   The   following features of an observed
galaxy may constrain the scope of this analysis.
\item{$\bullet$} {\sl  Thick disk, with random motions perpendicular 
to the plane of the galaxy.}

The measured line profile yields a mean value corresponding to
the integrated  emission through the width of the 
disk;
as the galaxy is not edge on, the result of the cross correlation of 
the spectra does not  give  $F_\phi$ and $F_R$ exactly, but rather gives the
projection onto the line of sight of the full velocity distributions $F_{\phi,z}(v_\phi,v_z,R)$ and 
$F_{R,z}(v_R,v_z,R)$. However, it is consistent with the thin disk approximation to assume 
that the motion in the plane is decoupled from that perpendicular to the plane.
The observed line profile, $F_{v}(R,v)$, then corresponds to the
 convolution of these functions with the component of the velocity distribution 
perpendicular to the plane of the galaxy, $F_z(R,v_z)$. Formally, for 
the line profile measured along the major axis, this reads 
$$ F_v(R,v) = \int_{-\infty}^{\infty} 
F_z[R,v \, \sin(i)+ u\, \cos(i) ] 
\, F_\phi[R,v \, \cos(i)- u\, sin(i) ] \, du \,  \EQN blabla  $$
where $v$ and $u$ are the  velocities along the line of sight, and
perpendicular to the line of sight respectively.
 A similar expression for the line profile measured along the minor axis involving
$F_R$ follows. Here, the angle $i$ measures the inclination of the
plane of the galaxy with
 respect to the line of sight.  At this level of approximation the
 velocity distribution normal to the plane is accurately described by
 a centered Gaussian distribution; its variance follows roughly from
 the equation of radial support and is used to deconvolve \Eq{blabla}
 (or practically to convolve the predicted line profile pairs and
 compare them to the data).  Note that a rough estimate of the error
 induced is $\sigma_z^2/\sigma_\phi^2 \tan^2(i)
\sim 1 \%  $ in our Galaxy when $i= 20$ degrees. 
Note that when looking at an ellipse that corresponds to a circle in
the galaxy's plane one would measure the dispersion 
$(\sigma_R^2+\sigma_\phi^2) \cos^2(i)/2 +(\sigma_R^2-\sigma_\phi^2)
\cos(2\phi) \cos^2(i)/2 + \sigma_z^2 \sin^2(i)$. If it is assumed
that the asymmetric drift hypothesis holds,   -- which can be checked by
measuring $\langle v_\phi \rangle$ -- then it should in principle be
possible to measure indirectly $\sigma_z$.
\item{ $\bullet$}{\sl Non-axisymmetric halo or disk, with dust and absorption.}

The departure from axisymmetry reduces the number of invariants --
angular momentum is not conserved,  which
invalidates this method.  Some fraction of the population of stars may then
follow chaotic orbits which cannot be characterised by a distribution function.
Some real galaxies  either present intrinsic
non axisymmetric features such as bars, spiral or lopsided arms.
Triaxial halos may induce warps within the disk, hence compromising  the
evaluation  of the kinematics.
Non-axisymmetrically distributed dust or molecular clouds may also affect  measurement
of the light from the stellar disk.  
On the positive side, non-axisymmetric
  features often appear as small perturbations which may
 only contribute weakly to the overall mass distribution.  Under such
circumstances, an estimate of the underlying distribution function for
the axisymmetric component of the disk may be extracted from the
data.
 
 Sets of initial conditions for an
N-body simulation may also be extracted from the distribution function. The
time evolution of the code would yield constraints on the stability of
the observed galaxy. The above analysis could be carried out on an
isolated S0 galaxy and on an Sb galaxy showing a grand design spiral
pattern (or alternatively  on an Sa presenting an apparent  lopsided
HI component to isolate plausible different instability modes). The
relative properties of the two galaxies and  their supposedly distinct
fates when evolved forward in time should provide  an unprecedented
link between the theory of galactic disks and the detailed observation
of the kinematics of these objects.
The authors\cite{Cannon} have written a numerical procedure to implement linear 
stability analysis on arbitrary mass models and kinematics for
 flat and round galactic disks.

\item{$\bullet$}{\sl  Data quality.}

The method described in section~2 involves a deconvolution of the
measured line profiles which yields the velocity distributions which, in
turn, are related by Abel transforms. 
Both transformations are very sensitive to the unavoidable noise
in the data. Sources of noise are poor seeing,   photon noise, detector noise,
thermal and mechanical drift of the spectrograph, poor telescope
tracking, etc...
 It is therefore desirable to construct from \Eqs{invertAbelR}, \Ep{FVR} and \Ep{formF} a set of pairs of Gaussian velocity distributions $(F_\phi,F_R)$ 
on which to project the observed quantities since this will help to assess the
errors induced by  this noise.
Indeed, \Eq{invertAbelR} is linear; therefore any superposition of Gaussian profiles 
corresponding to a good fit of the observed line profile will lead to
a  distribution function expressed in terms of sums of solutions of \Eq{invertAbelR}. 
 These Gaussians may in turn be identified with 
 populations at different temperatures -- say corresponding to a young
 component and 
an older population of stars.
Formally this translates as
$$ F_\phi(R,v_\phi)=\sum_i S_i(R)\,
\exp\left( -{ \left[ v_\phi- v_i(R) \right]^2 \over 2 \sigma_i^2(R)} \right) \, ,\EQN formFi $$ 
where the  functions $S_i(R)$,$\sigma_i(R)$ and $v_i(R)$  should be fitted to the observed
azimuthal line profiles.
Each component in \Eq{formFi} may then be identified with those in \Eq{formF2}, yielding the temperature
profile of that population.  
This approach should yield the best compromise between fitting the
radial and the azimuthal profile while avoiding the deconvolution of
noisy data. It also provides directly a least squares  estimate of the
errors.
The observations described in this section may be achieved
with today's technology.

\section{ Conclusions}
  Simple general inversion formulae to construct
 distribution functions for flat systems whose surface density and
 Toomre's Q number profiles were given  and illustrated.
 The purpose of these functions is to provide plausible galactic
 models and assess their critical stability with respect to global
 non-axisymmetric perturbations.  The inversion is carried out for a
 given azimuthal velocity distribution (or a given specific energy
 distribution) which may either be observed or chosen accordingly.
 When the azimuthal velocity distribution is measured from data, only
 a subset of the observationally available line profiles, namely the
 line profiles measured on the major axis, is required to re-derive
 the complete kinematics from which the line profile measured along
 the minor axis is predicted. This prediction may then be compared
with the observed minor axis line profile.
The Ansatz presented in section~3 yield direct and general
 methods for the construction of distribution functions compatible
 with a given surface density for the purpose of theoretical modelling.
 These distribution functions describe stable models with realistic
 velocity distributions for power law disks, and also the Isochrone
 and the Kuzmin disks. Note that the inversion technique 
described in section~2 can be implemented in the relativistic 
r\'egime as discussed by Pichon \& Lynden-Bell\cite{PLB2}.

\vskip 0.5cm
{\sl  CP wishes to thank
R. Cannon and A. Cooke for usefull conversations,
and the  anonymous referee for suggesting improvements 
in the organisation of the paper
}

%

\nosechead{References}
\ListReferences
\vfill
\eject

\bye